\documentstyle[nato]{crckapb} 

\begin{opening}
\subtitle{SPEC report 96/058 ( CEA Saclay ) \\ To appear in the Proceedings of the ``Magnetic Hysteresis in Novel Materials'' NATO Advanced Study Institute, Greece, July 1996 }

\title{Anomalies in the relaxation of small magnetic particles at very low temperatures}

\author{R. SAPPEY}
\author{E. VINCENT}
\author{J. HAMMANN}
\institute{e-mail : sappey@spec.saclay.cea.fr \\ Service de Physique de l'Etat Condens\'e, CEA Saclay\\
           91191 Gif sur Yvette Cedex, France}

\author{F. CHAPUT}
\author{J.P. Boilot}
\institute{Groupe de Chimie des Solides,\\  Laboratoire P.M.C., CNRS URA D1254,
\\Ecole Polytechnique, 91128 Palaiseau, France}

\author{D. Zins,}
\institute{Laboratoire de Physico-Chimie,\\ CNRS ER 44, Universit\'e Pierre et Marie Curie\\4 place Jussieu, 75252 Paris, France}

\end{opening}

\begin{document}


\input epsf
\epsfverbosetrue



\def\gam{$\gamma-Fe_2O_3$}

\def\magnetite{$\alpha Fe_2O_3$}

\begin{abstract}
 The magnetization relaxation rate of small \gam ~particles dispersed in a
silica matrix has been measured from  60 mK to 5 K. It shows a minimum around 150 mK, that can be discussed in terms of either thermal or quantum relaxation regime. 
\end{abstract}

\section{Introduction}
The magnetization dynamics of single-domain nanometric particles at low temperature is presently a subject of intense interest, in the hope of finding evidence for quantum tunneling of the magnetic moment through the anisotropy barrier associated with the particle \cite{chud}. Apart from some pioneering attempts at  a study of a unique particle \cite{werns}, most efforts are concentrated on macroscopic samples, in which an accurate knowledge of the effective distribution of barriers is difficult, hence hindering a clear interpretation of the results  \cite{barba1},\cite{T.Lnt}. Moreover, except in a few cases \cite{paulsen}, the low-temperature range of the published data is often limited to pumped-He cryogenic techniques ($\sim2K$), which still makes an unambiguous characterization of quantum effects more difficult.

In this paper, we present magnetic measurements which have been performed using a dilution refrigerator \cite{PP}, that allow data to be taken down to $\sim 50mK$. We have studied a sample of \gam ~particles, dispersed in a silica matrix, with a typical diameter of $\sim 6 \ nm$. The relaxation dynamics of \gam ~particles has already been shown to exhibit some anomalies \cite{Zhang}, that appear at the very end of the accessible temperature range (1.8 K). Our present data show that the relaxation rates in our sample do indeed fail to go down to zero when the temperature is lowered to $60 \ mK$.

\section{Sample characterization}

The small particles of \gam ~(maghemite) are embedded in a silica matrix, obtained
by a polymerization process at room temperature. They are diluted to a volume fraction of $4.10^{-4}$, in order to have them as independent as possible. The diameter distribution obtained  by transmission electron microscopy is shown in the inset of Fig. 1; it can be fitted to a log-normal shape with peak value $d_0=6.3\ nm$ and standard deviation $\sigma=0.25$.


\begin{figure}[h]

\centerline{\epsfysize=7cm \epsfbox{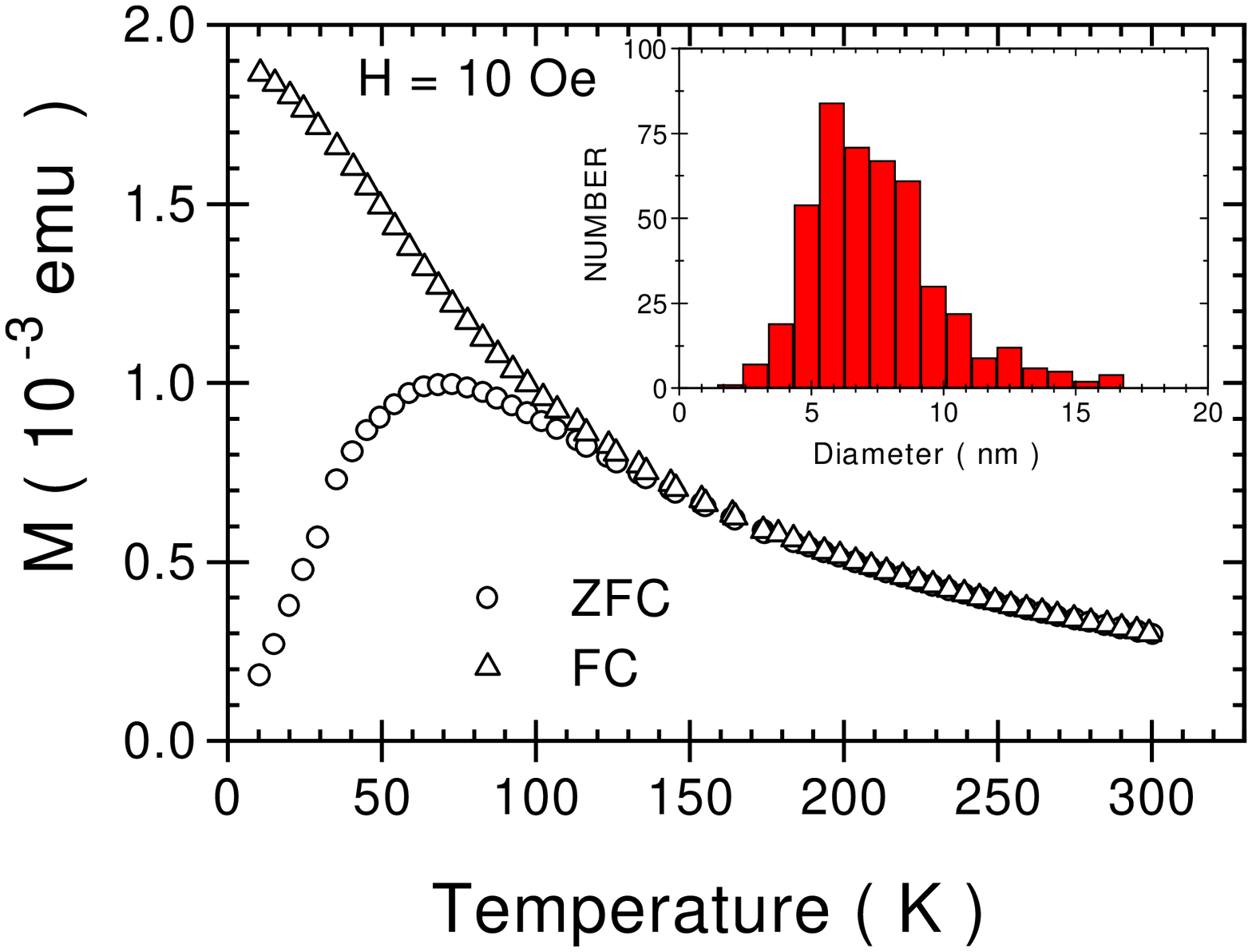} }

\caption{Total magnetic moment of the sample, measured in ZFC and FC procedures. The inset shows the size distribution of the particles deduced from transmission electron microscopy.}

\end{figure}


Fig.1 presents magnetic characterization measurements performed with a commercial SQUID magnetometer (Cryogenic Ltd). Here and all throughout the paper, we have plotted the measured magnetic moment in cgs units (corresponding to a total maghemite volume $\sim 4.10^{-5} \ cm^3$). The ``ZFC" curve is measured in the usual way by cooling the sample (down to 10 K) in zero field, applying a field and then raising the temperature;
 the field-cooled one (``FC") is obtained while cooling in the field $H$.

The ZFC curve shows a broad maximum around $T^{peak}\simeq 73 \ K$. It represents the progressive deblocking of larger and larger particles as the temperature T is raised. Let us consider that a particle of volume $V$ involves an anisotropy barrier $U=K_a.V$, where $K_a$ is a density of anisotropy energy.  If the time spent at a given T is $t$ ($\sim100\, s$), then for thermally activated dynamics most particles which are being deblocked at T have a typical volume V obeying an Arrhenius law 
\begin{equation}  \label{U=}
K_a.V=k_B.T.\ln{t\over \tau_0} \qquad ,
\end{equation}
where $\tau_0\sim 10^{-10}s$ is a microscopic attempt time \cite{dormann}. By assuming in addition that the saturated moment of a particle is proportional to its volume, and that the moments follow  a Langevin function when they are deblocked (super\-para\-magnetism), we have calculated the ZFC curve corresponding to the measured size distribution. The peak is obtained at the measured temperature for $K_a=7.5\, 10^5\,erg/cm^3$. This value is in agreement  \cite{romain} with high-field measurements where the integral of the work needed for saturating the sample has been evaluated and also with M\"ossbauer spectroscopy results. It is one order of magnitude larger than the bulk maghemite value, as commonly observed in small particles where shape and surface contributions have increased the magnetic anisotropy \cite{dormann}.
 
 Note that, due to the distribution width and to the 1/T variation of super\-para\-magnetism, the ZFC-peak  is found at a temperature which is three times larger than that corresponding to the peak value $d_0$ of the size distribution ($T_b(d_0)=25\, K$) \cite{romain}.

\section{Magnetic behavior towards very low temperatures} 

The setup used for the low-T experiments is a home made combination of an r.f. SQUID magnetometer \cite{ocio} and a dilution refrigerator \cite{PP}. The sample is coupled to the mixing chamber through a thermal impedance which allows a temperature range of 35 mK to 7 K. For relaxation measurements at the lowest temperatures, some spurious heating has been found when the field is varied, due to eddy currents in the thermalization link; we have therefore carefully adjusted the field amplitude, and chosen a ``slow'' cut-off procedure (5 s), in such a way that the results become independent of both these parameters. We also have limited our lower range to 60 mK.

The sample is first cooled in zero field from room temperature to the dilution regime. From that point, the temperature can no longer be easily raised above 7 K. The procedure for the relaxation measurements at $T_0\le 5\, K$ starts with heating the sample to a high enough temperature for deblocking of all particles which may participate in the dynamics at $T_0$, e.g. 7 K. Then the sample is field-cooled from 7 K to $T_0$, the field is decreased to zero and the SQUID signal variation corresponding to the slow relaxation processes is measured. This procedure of not heating up to room temperature makes sense because our sample is highly diluted; in a first approximation the particles can be considered independent of each other.  We have checked that our choice of the reinitialization temperature had no influence on the resulting dynamics.


\begin{figure}[h]

\centerline{\epsfysize=7cm \epsfbox{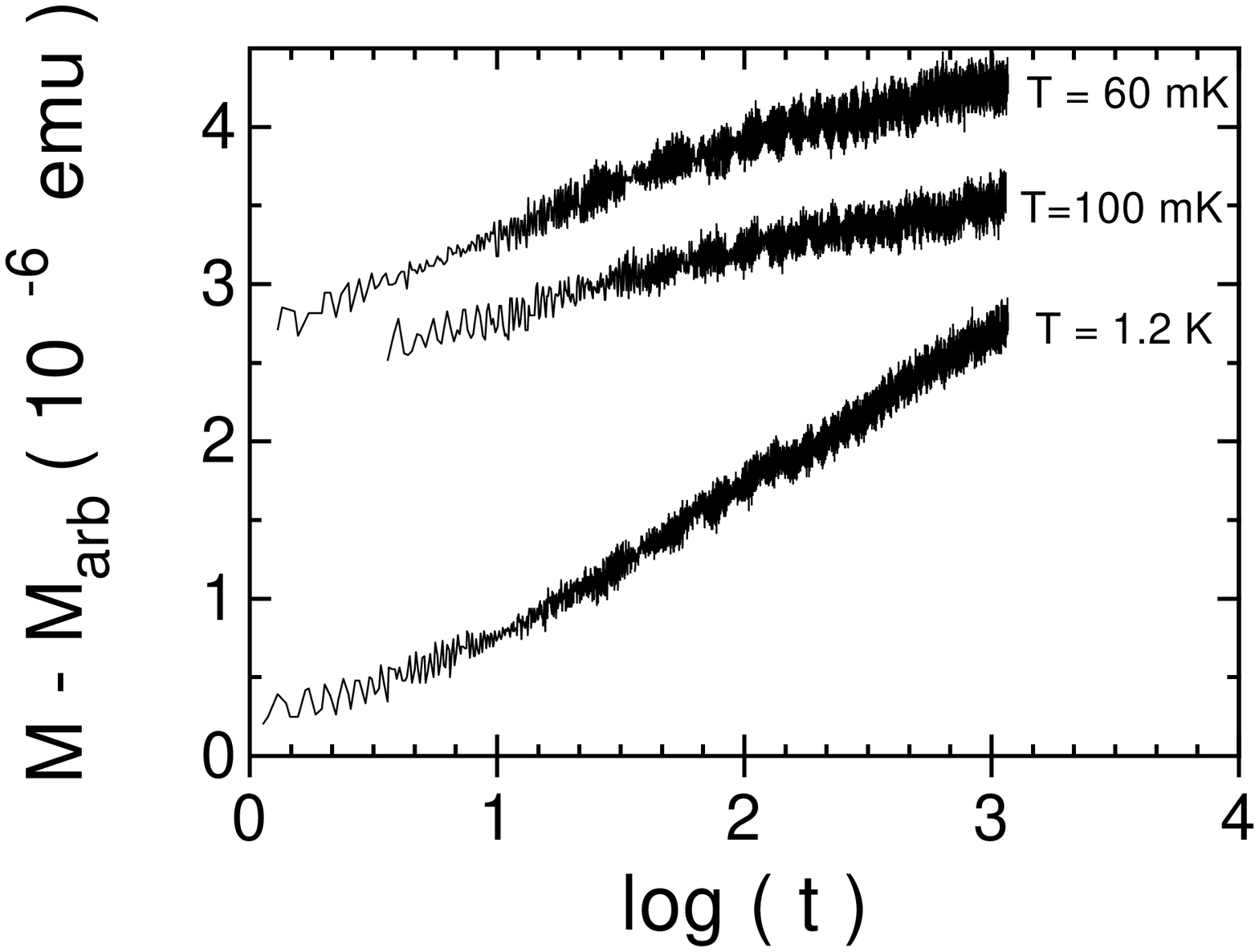} }

\caption{ Typical relaxation curves at low temperatures, as a function of the decimal logarithm of the time in seconds. The curves have been vertically shifted by arbitrary values. }

\end{figure}



\begin{figure}[h]

\centerline{\epsfysize=7cm \epsfbox{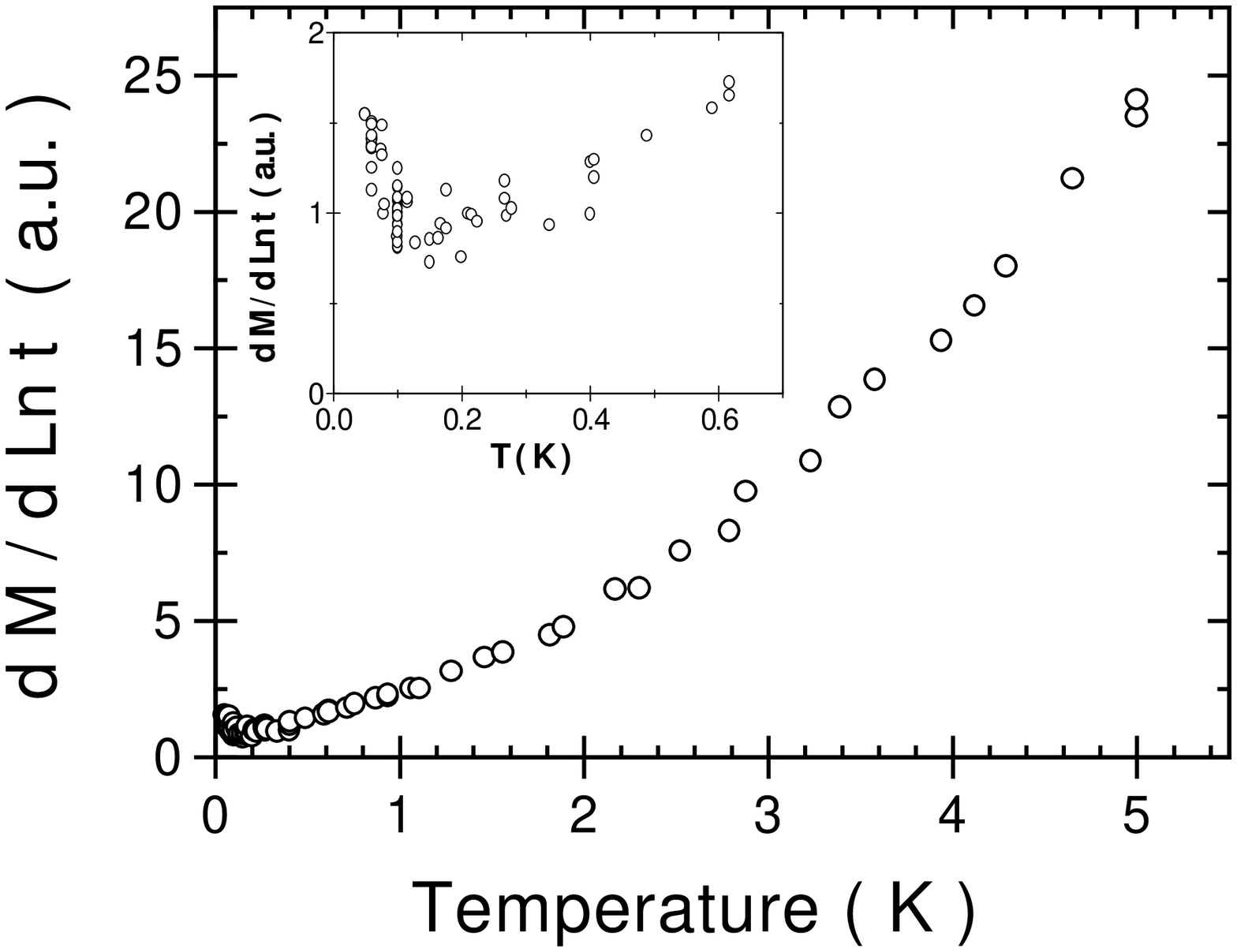} }

\caption{ Magnetic viscosity  as a function of temperature.}

\end{figure}


Figure 2 presents examples of relaxation curves. They are roughly logarithmic in time, apart from some uncertainty in the first seconds, which should be related to the 5 s field cut-off duration. In this paper, we only consider the average logarithmic slope of the curves (``magnetic viscosity"), which we determine between $10^2$ and $10^3\ s$. 

Figure 3 shows our set of results. For decreasing temperatures, the measured viscosity first decreases, then flattens out, and surprisingly increases back below 150 mK. We present a simple model for the T-dependence of the viscosity before discussing this result in more detail.

\section{A simple picture of thermal relaxation}

By thinking of the sample relaxation at T as a sum of independent processes, one may write the total relaxing moment $M_T(t)$ as
\begin{equation}    \label{M=sum} 
M_T(t)=\int_0^{+\infty  }\  m(U)\  P(U) \exp-{t\over \tau(U)}\ dU
\end{equation}
where the summation runs over the barrier distribution P(U) associated with the size distribution of the particles. $m(U)$ stands for the ``field-cooled moment" of the particles with anisotropy barrier U, which is the thermal average of the 
moments at their blocking temperature; as a first approximation, one may assume $U=K_a.V$ and $m(U)\propto V$, hence $m(U) \propto U$. At any temperature $T$ and after a time $t$ following the field cut-off, one may consider that the only relaxing objects are those for which $\tau(U)=t$. The logarithmic derivative $S$ of the magnetization (magnetic viscosity) can then be easily derived as
\begin{equation}   \label{S1}
S\equiv {\partial M_T \over \partial \ln t}\propto T.P(U_c).m(U_c) \quad\hbox{where} \quad U_c=k_B.T.\ln{t\over\tau_0} \quad . 
\end{equation}
The magnetic viscosity is commonly expected to be proportional to T \cite{street}, a controversial point
 since in our cases of interest the energy barrier distribution $P(U)$ may vary significantly \cite{barba1},\cite{T.Lnt}. Indeed, from Eq. \ref{S1}, one sees that the distribution of interest is $P(U).m(U)$ rather than $P(U)$ itself; with $m(U_c) \propto U_c$, Eq. \ref{S1} then becomes
\begin{equation} \label{S2}
S\propto T^2. \ln({t\over\tau_0}).\ P\Big( U_c=k_B.T.\ln{t\over \tau_0}\Big) \quad .
\end{equation}
We believe that these $t$ and $T^2$-dependences of the viscosity are probably hidden in most experimental results, due to the combination of the distributions $P(U)$ and $m(U)$ which are not accurately known (the $\ln^2 (t/\tau_0)$-variation of the magnetization is very close to $\ln t$, due to the microscopic value of $\tau_0$). However, it seems to us that the first approximation of the viscosity in the case of non-interacting particles with a flat distribution of barriers should be a quadratic rather than a linear function of temperature.

\section{Discussion}

As expected from thermally activated dynamics and a regular distribution of barriers, the 0.5-5 K viscosity is seen to decrease for decreasing temperatures. It shows a slight upwards curvature which is compatible with a $T^2$-dependence and a flat distribution; actually, 
 this T-range corresponds to the blocking of 2-3 nm objects, which are not well characterized from the distribution in Fig.1. However, it is clear from Fig.3 that a normal extrapolation will not yield a zero viscosity at zero temperature; below 150 mK, the viscosity data even show a systematic tendency to increase as T is lowered. A similar behavior has been noted in an array of cobalt particles \cite{wegrowe}, and also in a Permalloy sample \cite{Vitale}. With respect to maghemite, a viscosity anomaly (plateau from 2.2 to 1.8 K) has been observed in a system of particles dispersed in a glassy matrix \cite{Zhang}; no anomaly was visible for the same particles in water, suggesting the influence of the matrix via magneto\-striction phenomena \cite{Zhang}. 


\begin{figure}[h]

\centerline{\epsfysize=7cm \epsfbox{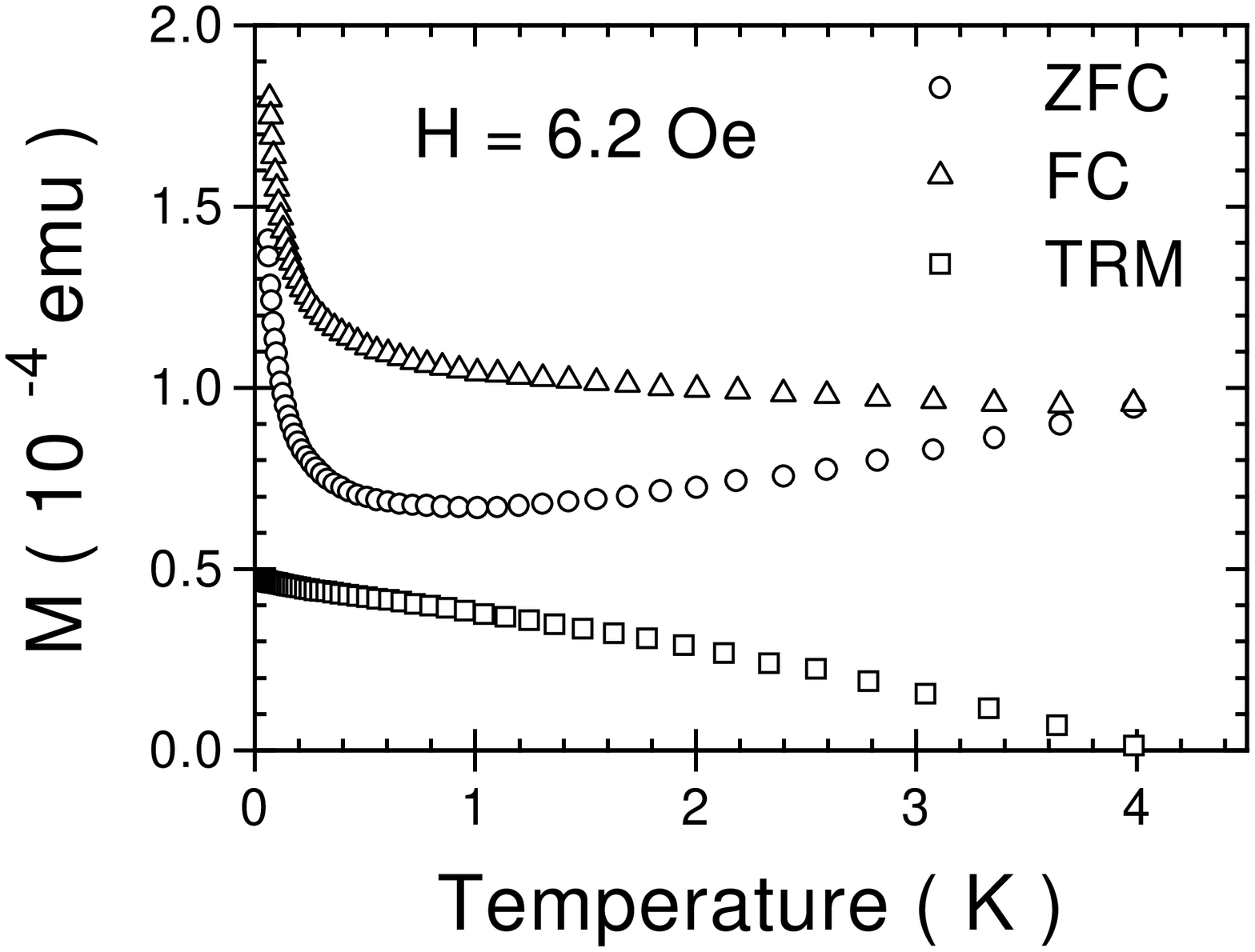} }

\caption{ ZFC and FC curves in the low temperature region. The ``TRM'' curve has been measured when, after field-cooling to 60 mK, the field is cut and the temperature is raised. This measured TRM is equal to the difference between FC and ZFC, as usual when linear response theory applies.}

\end{figure}


We consider that our present results may give rise to two possible conclusions (a combination of both is also possible).
First, one may assume that the dynamics is thermally activated. The implication of our results is that the distribution of energy barriers $P(U)$ increases abruptly towards  smaller values, more rapidly than $1/U^2$. This is a surprising result, very different from the framework in which viscosity measurements are commonly interpreted in the literature (approximately a flat distribution). We have in addition performed ZFC/FC measurements in this low-T range, which are displayed in Fig.4. They show an increase in the magnetization for decreasing T, which is 1/T-like and of the same amount in both ZFC and FC cases (see TRM in Fig. 4). If this behavior is ascribed to clusters of e.g. 10 spins, the Curie constant would correspond to 0.5\% of the total \gam ~amount. Thus, there are indeed some very small magnetic entities which are not frozen, even at 60 mK. Fig. 4 also shows a significant difference between the ZFC and FC curves, which corresponds to the slow dynamics observed in this low-T range. All data are therefore compatible with the existence of a significant low-energy tail of the barrier  distribution, increasing further for the lowest values. One may think of very small particles; it would be of interest to check other systems of small particles for this possibility. It has also been proposed that such small barriers arise from decompensation effects at the surface of the ferrimagnetic particles \cite{berko}; surface defects might be an intrinsic component of the dynamics of nanometric particles at very low temperatures.

A second possible conclusion concerns the quantum tunneling of the particle magnetization (QTM) through its anisotropy barrier. 
In a first approximation, the contribution of such processes could be independent of temperature; from \cite{chud}, quantum processes should be of the same order of magnitude as thermal processes below a crossover temperature $T_c$, which can be here estimated as $T_c\simeq 100\,mK$ ($T_c$ does not depend on the barrier height, which only influences the relaxation rates). It is therefore possible that such processes contribute significantly in our T-range (one may even wonder why they should not be visible). The increase of the viscosity towards lower T can be understood in two ways. On the one hand, it has been argued in \cite{Vitale} that the viscosity should be T-independent if the two energy levels between which quantum tunneling occurs are sufficiently separated with respect to $k_BT$, whereas it should go like 1/T for quasi-degenerate levels, which could be our situation of low-field relaxations. A low-T increase of the viscosity in Permalloy has thus been described as quantum jumps of a Bloch wall between pinning sites of comparable energies \cite{Vitale}. In more general terms, on the other hand, one may think that lower temperatures decrease the coupling to phonons, therefore reducing the dissipation and enhancing quantum tunneling processes \cite{chud2}.

A ``T.Lnt" plot has been proposed to help distinguish between thermal and quantum processes in size-distributed particles \cite{T.Lnt}, but this is not possible with the present relaxation data, obtained by measuring only SQUID signal variation (and not the full value of the magnetization). Actually, the question of a satisfactory evidence of QTM processes in such systems remains controversial; however, we believe that the numerous observations of anomalies in the low-T dynamics of small particles lead us to the minimal conclusion that things are not as simple as we had thought.


\begin{thebibliography}{}
\bibitem{ocio}
The SQUID magnetometry has been designed and patiently tuned by M. Ocio, whom we also thank for stimulating scientific discussions.
\bibitem{PP}
The special-purpose dilution refrigerator has been designed and built in CEA-Saclay by P. Pari and Ph. Forget, with the help of L. Le Pape.
\end{thebibliography}
\end{document}